\documentclass[preprint,aps,pra,showpacs,floatfix,superscriptaddress]{revtex4-1}
\usepackage[T1]{fontenc}
\usepackage{longtable}
\usepackage[referable]{threeparttablex}

\usepackage{graphicx}
\usepackage{amsmath,amsfonts,amssymb,verbatim,float}

\usepackage[usenames]{color}	%for colours
\usepackage{colortbl}
\usepackage{calrsfs}
\usepackage{ulem} % for remarks

\usepackage{hyperref} % for links in the bibliography
\usepackage{subfigure}
\usepackage{indentfirst}
\usepackage{siunitx}
\definecolor{faded}{gray}{0.45}
\renewcommand{\vec}[1]{\mathbf{#1}}
\newcommand{\balpha}{\boldsymbol{\alpha}}

\begin{document}

\thispagestyle{empty}
\title{
Relativistic Stark energies of hydrogenlike ions
}
\author{I.~A.~Maltsev}
\affiliation{
Department of Physics, St. Petersburg State University,
Universitetskaya Naberezhnaya 7/9, 199034 St. Petersburg, Russia
}
\author{D.~A.~Tumakov}
\affiliation{
Department of Physics, St. Petersburg State University,
Universitetskaya Naberezhnaya 7/9, 199034 St. Petersburg, Russia
}
\author{R.~V.~Popov}
\affiliation{
Petersburg Nuclear Physics Institute named by 
B.~P.~Konstantinov of National Research Centre
“Kurchatov Institute”, Gatchina, Leningrad District 188300, Russia
}
\affiliation{
Department of Physics, St. Petersburg State University,
Universitetskaya Naberezhnaya 7/9, 199034 St. Petersburg, Russia
}
\author{V.~M.~Shabaev}
\affiliation{
Department of Physics, St. Petersburg State University,
Universitetskaya Naberezhnaya 7/9, 199034 St. Petersburg, Russia
}
\affiliation{
Petersburg Nuclear Physics Institute named by 
B.~P.~Konstantinov of National Research Centre
“Kurchatov Institute”, Gatchina, Leningrad District 188300, Russia
}
%
% BEGIN ABSTRACT ==============================================================
%
\begin{abstract}
The relativistic energies and widths of hydrogenlike ions exposed to 
the uniform electric field
are calculated. The calculations are performed for the ground and lowest
excited states using the complex scaling technique in combination 
with a finite-basis method. The obtained results are compared with the
nonrelativistic values. The role of relativistic effects is investigated.
\end{abstract}
%
% END ABSTRACT ================================================================
%
\maketitle
%
% BEGIN INTRODUCTION ==========================================================
%
% Plan ------------------------------------------------------------------------
%
%
% -----------------------------------------------------------------------------
%
\section{INTRODUCTION}
The bound states of an atom placed in a uniform electric field are shifted and turn into 
resonances. The resonance states are embedded into the continuum and have
finite energy width. This means that the atomic electrons can escape via 
tunneling through the potential barrier formed by the Coulomb and uniform electric
fields. This phenomenon is referred to as a Stark effect and for many years
has been studied in atomic systems 
experimentally~\cite{Traubenberg_81, Stebbings_76, Littman_76, Koch_81, Bergman_84, Stodolna_13} 
as well as
theoretically~\cite{Hehenberger_74, Zapryagaev_78, Benassi_80, Farreley_83, Gallas_82, Damburg_76, Damburg_78, Maquet_83, Kolosov_87, Lai_81, Kolosov_83, Lin_11, Rao_94, Fernandez_96, Jenschura_01, Milosevic_02, Milosevic_02_2, Popov_04, Ivanov_04, Batishchev_10, Ferandez-Menchero_13, Rozenbaum_14, Fernandez_18, Maltsev_21}. 
Many theoretical approaches have been applied for calculation of Stark resonances. 
However, almost all of these calculations were nonrelativistic. The relativistic 
effects can have some impact even in light systems (see Ref.~\cite{Ivanov_04}).
The relativistic treatment is required for searching for parity-nonconserving (PNC)
effects and physics beyond the standard
model in molecules, where the Stark shifts play an important role (see, e.g., Refs.~\cite{Andreev_18, Blanchard_23}). 
For heavy ions the relativistic consideration is absolutely necessary. 
Meanwhile, experiments with heavy partially stripped ions (PSIs) in very strong electric fields 
will soon become feasible.

One of the new projects, proposed currently as a part of the Physics Beyond Colliders initiative, 
is the Gamma Factory~\cite{Budker_20}. The proposed idea
is to combine the relativistic beams of heavy PSIs
at the Large Hadron Collider with the laser facility and use the Doppler boosting 
of the laser photons in the PSI reference frame. 
The PSI spectroscopy in strong external fields is one of the promising research 
topics of the project. If the PSI beam placed
in the transverse magnetic field, then in the PSI rest frame there exists an electric 
field enhanced by the $\gamma$ factor.
Modern high-field magnets allow generation 
of electric fields in the PSI rest frame of 
strength up to $10^{12}$~V/cm or even higher~\cite{Budker_20}.
A field of such strength allows manipulating the
energy levels of heavy PSIs. The theoretical values of resonance positions 
seem to be highly required for such investigations. The values of the Stark widths 
are also important for estimation of ion beam stability, since ion losses
due the Stark ionization of PSI can take place. 

For relatively weak fields the positions of the Stark resonances can be 
calculated using the relativistic 
perturbation theory~\cite{Zapryagaev_78, Rozenbaum_14}. 
In Ref.~\cite{Rozenbaum_14}, the relativistic resonance positions were 
also obtained via numerically solving the Dirac equation in
a finite basis set, which allows us to take into account
the external field exactly. However, since the resonance wave functions are not square integrable, 
the standard Hermite finite-basis-set methods
cannot provide accurate values of the resonance positions~\cite{Zaytsev_19}. 
Moreover, they cannot be directly used for calculation of the resonance widths. 
The relativistic values of the resonance widths 
were obtained in Refs.~\cite{Milosevic_02, Milosevic_02_2, Popov_04} using the semiclassical approximation. 
The semiclassical approach allows us to obtain the corresponding analytical expressions, 
but its accuracy is limited. 

The precise values of the resonance positions as well as the resonance widths can be calculated 
with the complex-scaling (CS) method. The CS technique is based on dilation of the Hamiltonian 
into the complex plane. After the dilation the resonances appear as square-integrable solutions of the 
Dirac equation. The corresponding energies have complex values. The real part of the
complex energy matches the resonance position and the imaginary part defines the 
resonance width. Previously, the CS method was successfully employed for 
relativistic calculations of many-electron 
autoionization states~\cite{Kieslich_04, Mueller_18, Zaytsev_19, Zaytsev_20}
and supercritical resonance 
in heavy quasimolecules~\cite{Ackad_07_1, Ackad_07_2, Marsman_11, Maltsev_20}.
Recently, the CS method was implemented in Q-Chem quantum chemistry program package, 
which is also able to take into account some relativistic effects~\cite{Jagau_18, Epifanovsky_21}.  
A detailed description of the complex-scaling approach and 
its applications can be found 
in reviews in~\cite{Reinhardt_ARPC33_223:1982, Junker_AAMP18_107:1982, Ho_PRC99_1:1983, Moiseyev_PR302_211:1998, Lindroth_AQC63_247:2012}.

The relativistic CS method was used previously for calculations of Stark energies and widths
of one-electron atomic systems
in Refs.~\cite{Ivanov_04, Maltsev_21}. 
However, the calculations were restricted 
to only hydrogen and hydrogenlike neon. 
%It should be noted that the nonrelativistic Stark energies for a hydrogenlike 
%ion with the nuclear charge $Z$ in the pointlike nuclear model can be
%easily obtained from the hydrogen values by multiplying the field strength by $Z^3$, 
%and energies and widths on $Z^2$. But there is no such scaling law in the relativistic 
%case.
It should be noted that   
the Stark energies and widths for a hydrogenlike 
ion with the nuclear charge $Z$ exposed to an electric field $F$ can be
easily obtained from the corresponding hydrogen values calculated 
for the field strength $F/Z^3$ by multiplying them by $Z^2$.
This scaling law is a direct consequence of the Schr\"odinger equation 
with a pointlike nucleus and there is no such rule for the relativistic case. 
Therefore, the relativistic calculations should be performed for every $Z$ under 
consideration. Taking into account the finite nuclear size also breaks the scaling law.

The aim of the present work is to fill the gap in theoretical data and 
investigate the influence of the relativistic effects on the Stark resonances. 
In order to achieve this aim we have performed the calculations of the lowest resonance
states for several hydrogenlike ions between $Z = 1$ and $82$. 
The resonance parameters are obtained utilizing the relativistic CS technique.
After the complex scaling, the Dirac equation is solved using the finite-basis method 
described in Refs.~\cite{Maltsev_18, Maltsev_17}. 
The obtained results are compared with available 
nonrelativistic and relativistic values and the 
influence of the relativistic effects is investigated. 

Throughout the paper we assume $\hbar = 1$.

\section{THEORY}
%
% - - - - - - - - - - - - - - - - - - - - - - - - - - - - - - - - - - - - - - -
%
The relativistic energy spectrum of a hydrogenlike ion is determined by the Dirac equation
\begin{equation}
    H \psi(\vec{r}) = E \psi(\vec r),
    \label{eq:dirac}
\end{equation}
where, in the presence of an external uniform electric field, the Hamiltonian has the following form: 
\begin{equation}
    H = c (\vec{\balpha} \cdot \vec{p}) + V_{\rm nucl} (r) + eFz + \beta m_e c^2.
\label{eq:ham_dirac}
\end{equation}
Here $e$ is the electron charge ($e = |e|$),
$V_{\rm nucl} (r)$ is the nuclear potential,
and $F$ is the strength of the electric field which is assumed to be directed 
along the $z$ axis. For the nuclear potential, the pointlike nuclear model 
($V_{\rm nucl} (r) = -eZ / r$) is generally 
used. However, in many cases, especially for heavy ions, the finite-nuclear-size 
effect is rather significant.
Therefore, in the present work we utilize the model of a uniformly charged sphere,
which takes into account the finite nuclear size:
\begin{equation}
    V_{\rm nucl}(r) = 
    \left\lbrace
    \begin{aligned}
        & -\frac{eZ}{2R_{\rm nucl}}
        \left(3 - \frac{r^2}{R_{\rm nucl}^2}\right),
        & r < R_{\rm nucl}\\
        & -\frac{eZ}{r},
        & r > R_{\rm nucl},
\end{aligned}
\right.
\label{eq:pot_sphere}
\end{equation}
where $R_{\rm nucl} = \sqrt{5/3} R_{\rm RMS}$ is the nuclear radius and 
$R_{\rm RMS}$ is the root-mean-square nuclear radius. 

The Dirac equation is considered in the spherical coordinate system $(r, \theta, \varphi)$.
The Hamiltonian~\eqref{eq:ham_dirac} is invariant under rotation around the $z$ axis.
Therefore, it is possible to separate the azimuthal angle~$\varphi$ from other coordinates.
The separation can be done by substitution of the function   
\begin{equation}
  \psi_m(r, \theta, \varphi)=
 \frac{1}{r}
 \begin{pmatrix}
    G_1(r, \theta) \exp [i(m-\frac{1}{2})\varphi] \\
    G_2 (r, \theta)\exp [i(m+\frac{1}{2})\varphi] \\
   i F_1 (r, \theta)\exp[i(m-\frac{1}{2})\varphi]\\
   i F_2 (r, \theta)\exp[i(m+\frac{1}{2})\varphi]
  \end{pmatrix}
  \label{eq:m_function}
\end{equation}
into the Dirac equation~\eqref{eq:dirac}. Here $m$ is the half-integer $z$ projection of the total angular
momentum. With this substation, Eq.~\eqref{eq:dirac} can be reduced to the following form:
\begin{equation}
  H_m  \, \Phi(r, \theta) = E \Phi(r, \theta).
  \label{eq:2d_dirac}
\end{equation}%
Here the four-component wave function $\Phi(r, \theta)$ is given by
\begin{equation}
   \Phi(r, \theta) = 
   \begin{pmatrix}
    G_1(r, \theta)\\
    G_2(r, \theta)\\
    F_1(r, \theta)\\
    F_2(r, \theta)
 \end{pmatrix}
 \label{eq:rth_fun}
\end{equation}
and the Hamiltonian $H_m$ can be represented as
\begin{equation}
   H_{m} (t) =
    \begin{pmatrix} \displaystyle
      m_e c^2 + V_{\rm nucl} (r) + eFz \quad & c \, D_m \\
      - c \, D_m & -m_e c^2 + V_{\rm nucl} (r) + eFz
    \end{pmatrix} ,
    \label{eq:2d_ham}
\end{equation}
\begin{equation}
\begin{split}
 D_m = \left( \sigma_z \cos \theta + \sigma_x \sin \theta \right)
 \left( \frac{\partial}{\partial r} - \frac{1}{r} \right)\\
 + \frac{1}{r} \left(\sigma_x \cos \theta - \sigma_z \sin \theta \right) 
 \frac{\partial}{\partial \theta} \\
 + \frac{1}{r \sin\theta} \left( i m \sigma_y + \frac{1}{2} \sigma_x \right),
\end{split}
\end{equation}
where $\sigma_x$, $\sigma_y$, and $\sigma_z$ are the Pauli matrices.

Due to the presence of the uniform field $F$ Eqs.~\eqref{eq:dirac} and~\eqref{eq:2d_dirac} have
no bound states. For nonzero $F$ the original (at $F = 0$) bound states of a hydrogenlike 
ion become embedded in the positive continuum and can be described as resonances. 
The resonances have finite energy widths~$\Gamma$, which correspond to the probability of 
the electron being ionized via escaping through the potential barrier. 
In order to obtain the resonance positions and widths, we used the CS method. 
The simplest version of the CS technique is the uniform complex rotation,
according to which the radial coordinate is transformed as $r \rightarrow re^{i\Theta}$, where
$\Theta$ is a constant angle of the complex rotation. 
For the potential of a pointlike nucleus this transformation 
causes no problem and can be easily performed. However, if the potential
is not an analytic function, then the uniform complex rotation can not be done. 
In particular, the potential of the uniformly charged sphere given by Eq.~\eqref{eq:pot_sphere} is not analytic.
In order to overcome this obstacle, one can use the exterior complex scaling (ECS)
proposed in Ref.~\cite{Simon_79}:
\begin{equation}
    r \rightarrow 
    \left\lbrace
    \begin{aligned}
        & r, &  r \leqslant r_0,\\
        & r_0 + (r - r_0)e^{i \Theta}, & r > r_0.
     \end{aligned}
     \right.
    \label{eq:ecs}
\end{equation}
By such a transformation the internal region $r \leqslant r_0$
remains untouched while the complex rotation is performed in the external region, 
where the potential is analytic. The drawback, however, 
is that after the substitution~\eqref{eq:ecs} the derivative of 
the Hamiltonian eigenfunction is discontinuous at $r = r_0$. 
Therefore, in order to get a correct finite-basis representation
of the Dirac equation, one should use the basis functions which 
are also discontinuous at this point. 
Instead, in the present work we use a more universal version of the CS technique, 
namely, the smooth exterior complex scaling~\cite{Moiseyev_88, Moiseyev_PR302_211:1998}, 
which is defined by the transformation
\begin{equation}
     r \rightarrow 
    \left\lbrace
    \begin{aligned}
        & r, &  r \leqslant r_0,\\
        & r + (r - r_0) \left( e^{i \Theta} -1 \right) f(r), & r > r_0,
     \end{aligned}
     \right.
    \label{eq:sm_ecs}
\end{equation}
where the function $f(r)$ was chosen as
\begin{equation}
 f(r) = 1 - e^{-\left(\frac{r - r_0}{a}\right)^2}.
 \label{eq:cs_contour}
\end{equation}
This transformation defines a smooth transition from $r$ to $re^{i \Theta}$ 
for $r \rightarrow \infty$.
It worth mentioning that there also exists a complex absorbing potential (CAP) approach,
which is quite close to the smooth ECS method~\cite{Riss_93}. 
A similar complex-scaling contour $f(r)$ was used in Ref.~\cite{Elander_98}.
The parameters $r_0$ and $a$ can be 
adjusted in order to facilitate the convergence of the numerical calculation.
The smooth ECS is more flexible than the "sharp" one defined by Eq.~\eqref{eq:ecs}.
It should be noted, however, 
that, at least in some cases, the "sharp" ECS 
can provide more stable results than its smooth counterpart~\cite{Pirkola_22}.

After the transformation~\eqref{eq:ecs} or~\eqref{eq:sm_ecs} 
the Stark resonances match the square-integrable solutions 
of Eq.~\eqref{eq:2d_dirac} and the corresponding energy $E$ has a complex value:
\begin{equation}
    E = E_0 - i \Gamma / 2.
    \label{eq:comp_e}
\end{equation}
The real part $E_0$ is the position of the resonance, and the imaginary part defines the resonance width~$\Gamma$.

The complex-rotated Dirac equation is solved using the finite-basis method. The wave function 
$\Phi(r, \theta)$ (see Eq.~\eqref{eq:m_function}) is expanded as
\begin{equation}
 \Phi(r, \theta) = \sum\limits_{n = 1}^N C_n W_n (r, \theta).
 \label{eq:fin_exp}   
\end{equation}
The basis functions~$W_n(r, \theta)$ are constructed from $N_{\theta}$
B-splines dependent on the $\theta$ coordinate and $N_r$ B-splines dependent on the $r$
coordinate. The total number of basis functions is $N = 4 \times N_r \times N_{\theta}$. 
The construction is performed using the dual-kinetic balance~(DKB) technique for axially 
symmetric systems. This technique prevents the appearance of spurious states in the spectrum.
A detailed description of the employed basis set can be found in Ref.~\cite{Maltsev_18}. 
By the substitution of Eq.~\eqref{eq:fin_exp}, the Dirac equation~\eqref{eq:2d_dirac} is reduced 
to the generalized eigenvalue problem
\begin{equation}
 \sum\limits_{k = 1}^N H_{jk} C_k=\sum\limits_{k = 1}^N E S_{jk} C_k \, .
 \label{eq:gen_eigen}
\end{equation}
Here $H_{jk}$ and $S_{jk}$ correspond to the Hamiltonian and overlap matrices, respectively. 
The complex eigenvalues $E$ are found using the numerical diagonalization procedure. 
\section{RESULTS AND DISCUSSIONS}
In the present work, only states with the projection of total angular momentum
$m = 1/2$ are considered. The complex Stark energies are obtained by solving 
the eigenvalue problem~\eqref{eq:gen_eigen}. The resonance positions and widths 
are related to the complex eigenvalues via Eq.~\eqref{eq:comp_e}. 

For each nuclear charge $Z$ considered
the basis set is constructed from the B-splines defined in a box of 
size $r_{\rm max} \approx 174 / Z$~a.u. The radial B-spline knots are distributed
uniformly inside the nucleus and exponentially outside. 
%uniformly for $r\leqslant R_{\rm nucl}$ and exponentially for $R_{\rm nucl} < r < r_{\rm max}$.
We use the smooth ECS technique with the contour defined by Eq.~\eqref{eq:cs_contour} 
with $a = 7.75 / Z$~a.u. 
The following values of the contour parameter $r_0 / Z$ are chosen 
depending on the electric field strength $F$ and the atomic state under consideration:
$11$, $8$, and $5$ for the ground state, with $F /Z^3 \leqslant 0.04$, 
$0.04 < F/Z^3 \leqslant 0.07$, and $F/Z^3 > 0.07$, respectively, and $5$ for all 
excited states (all quantities are given in atomic units). By adjusting the values of 
$r_0$ and $a$ it is possible to improve the stability and convergence 
of the energy values. Note, however, that
accurate results can be obtained with a quite broad range of these parameters.   

\begin{figure}[ht]
    \centering
    \includegraphics{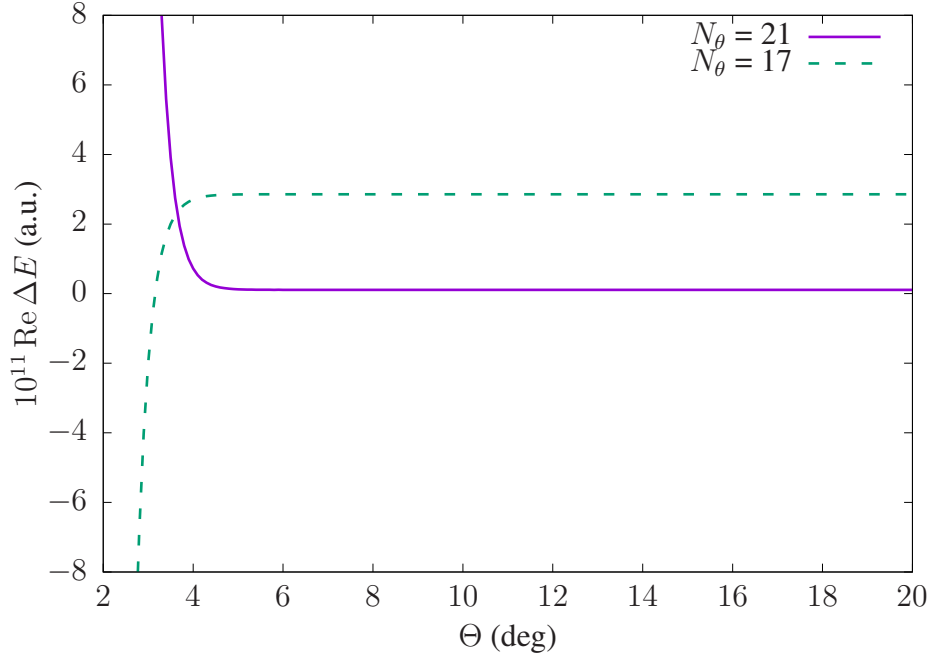}
    \caption{Real part of the complex Stark energy~$E$ 
    as a function of the complex-scaling angle~$\Theta$ for a hydrogen atom
    exposed to an electric field of strength $F = 0.05$~a.u. 
    Here $\Delta E = E - E_0$, where ${\rm Re} \, E_0 = -0.5061117144$~a.u.
    The solid line corresponds to the results obtained using the basis set with 
    $N_{\theta} = 21$ and the dashed line corresponds to the values obtained with $N_{\theta} = 17$.
    }
    \label{fig:real_theta}
\end{figure}

\begin{figure}[ht]
    \centering
    \includegraphics{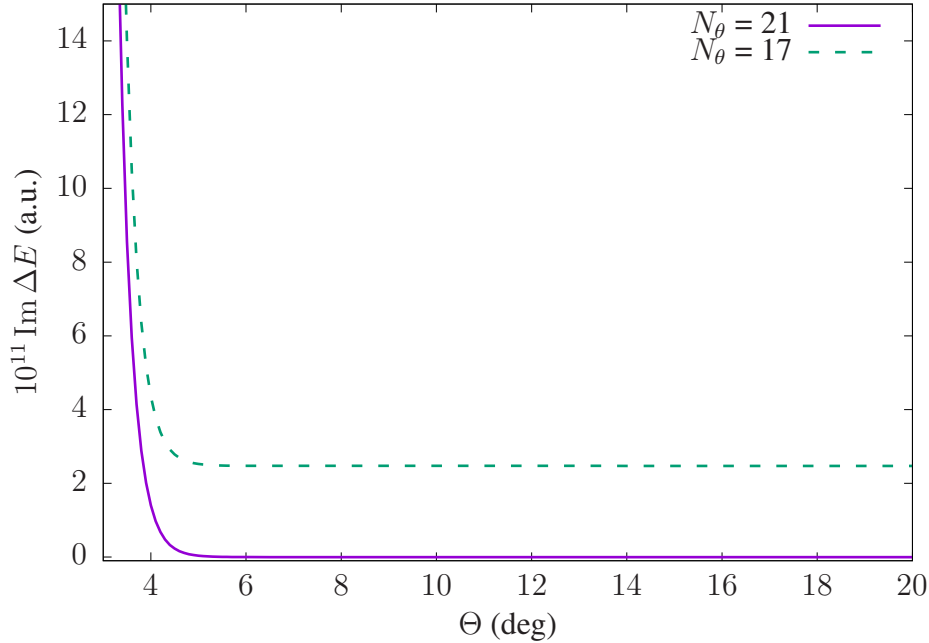}
    \caption{Imaginary part of the complex Stark energy~$E$ 
    as a function of the complex-scaling angle~$\Theta$ for hydrogen atom
    exposed to an electric field of strength $F = 0.05$~a.u. 
    Here $\Delta E = E - E_0$, where ${\rm Im} \, E_0 = -3.85812927 \times 10^{-5}$~a.u.
    The solid line corresponds to the results obtained using the basis set with 
    $N_{\theta} = 21$ and the dashed line corresponds to the values obtained with $N_{\theta} = 17$.
    }
    \label{fig:imag_theta}
\end{figure}

The exact solutions of the complex-scaled Dirac equation corresponding to the resonances  
do not depend on the angle of complex scaling~$\Theta$. However, the solutions of the finite-basis
representation~\eqref{eq:gen_eigen} exhibit such a dependence. 
In our case, the rapid change in real and imaginary part of 
the complex energy $E$ for small values of $\Theta$ 
is followed by a long plateau (see Figs.~\ref{fig:real_theta}
and~\ref{fig:imag_theta} for the real and imaginary parts, respectively).
Despite the fact that the energy values are not perfectly stable on the plateau, 
as can be seen from Figs.~\ref{fig:real_theta} and~\ref{fig:imag_theta}, 
the dependence on $\Theta$ is much smaller than the difference 
between the values obtained with the basis sets of close sizes.
This shows that the $\Theta$ dependence is negligible in comparison with the 
uncertainty which comes from the basis convergence.

The calculations are performed using the basis sets 
of different sizes for a wide range of CS angle $\Theta$. 
As an example, in Table~\ref{tab:conv} 
we present the results for a hydrogen atom
exposed to an electric field $F = 0.05$~a.u., 
which are obtained utilizing different numbers 
of radial and angular B-splines ($N_r$ and $N_\theta$, respectively). 
The largest employed basis set had the following parameters: $N_r = 500$ 
and $N_{\theta} = 21$ with the total number of the basis functions $N = 40000$.
The calculation uncertainty is estimated from the convergence of the results.
The estimation is done in such a way that the estimated uncertainty is 
well above the difference between the value for the largest basis set and 
any reasonable interpolation of the results to the complete-basis-set limit.
\begin{table}[ht]
 \caption{Real~$E$ and imaginary~$-\Gamma/2$ parts of the complex energy 
  of a hydrogen atom exposed to a uniform electric field $F = 0.05$~a.u. 
  The calculations were performed with the basis set constructed using $N_r$ radial and 
  $N_\theta$ angular B-splines.}
 \label{tab:conv}
 \begin{tabular}{l|
                S[exponent-mode = fixed, fixed-exponent = 0, round-mode = places, round-precision = 12, table-format= -1.12]
                S[exponent-mode = fixed, fixed-exponent = 0, round-mode = places, round-precision = 12, table-format= -1.12] 
                S[exponent-mode = fixed, fixed-exponent = 0, round-mode = places, round-precision = 12, table-format= -1.12]|
                S[round-mode = places, round-precision = 8, table-format= -1.8]
                S[round-mode = places, round-precision = 8, table-format= -1.8]
                S[round-mode = places, round-precision = 8, table-format= -1.8]}
\hline\hline
$N_{\theta}$ & \multicolumn{3}{c}{$E$ [a.u.]}       & \multicolumn{3}{c}{$ - \Gamma / 2 \times 10^5$ [a.u.]}              \\
              & {$N_r = 300$} & {$N_r = 400$} & {$N_r = 500$} & {$N_r = 300$} & {$N_r = 400$} & {$N_r = 500$} \\
\hline
11 & -5.061117080530495e-01  & -5.061117080619328e-01  & -5.061117080642072e-01  & -3.857925249956013e-05  & -3.857924803664866e-05  & -3.857924579481499e-05  \\
13 & -5.061117142380951e-01  & -5.061117142427614e-01  & -5.061117142431576e-01  & -3.858125702751084e-05  & -3.858125755657403e-05  & -3.858125762709890e-05  \\
15 & -5.061117143218867e-01  & -5.061117143272358e-01  & -5.061117143280267e-01  & -3.858120795996053e-05  & -3.858120839814137e-05  & -3.858120846395354e-05  \\
17 & -5.061117143652148e-01  & -5.061117143706948e-01  & -5.061117143714385e-01  & -3.858126758351663e-05  & -3.858126793771403e-05  & -3.858126795752455e-05  \\
19 & -5.061117143848235e-01  & -5.061117143902913e-01  & -5.061117143910498e-01  & -3.858128555180535e-05  & -3.858128591224404e-05  & -3.858128593260892e-05  \\
21 & -5.061117143926915e-01  & -5.061117143981432e-01  & -5.061117143989050e-01  & -3.858129233601658e-05  & -3.858129269257646e-05  & -3.858129271526616e-05  \\
\hline\hline

\end{tabular}

\end{table}

In order to investigate the impact of the relativistic effects on the Stark resonances,
we perform calculations for several hydrogenlike ions between $Z = 1$ and $82$.
The obtained results are compared with the corresponding nonrelativistic values. The latter
ones can be trivially obtained (in the pointlike nuclear model) for every $Z$ 
from the hydrogen values using the scaling law $F \rightarrow Z^3 F$, $E_0 \rightarrow Z^2 E_0$,
and $\Gamma \rightarrow Z^2 \Gamma$, where $F$ is the field strength, $E_0$ and $\Gamma$ 
are the resonance position and width, respectively. 
Here and below the "relativistic effects" refer
to the differences between the solutions of the Dirac and Schr\"odinger equations. 
They naturally include all the relativistic corrections (such as spin-orbit correction), 
which are usually used to improve the accuracy of nonrelativistic values.  
It should be noted, however, that in our calculations the finite nuclear model is utilized, while 
the scaled nonrelativistic values imply the pointlike nucleus. 
But we found that the finite-nuclear-size contribution is relatively small and does not qualitatively affect
the results. 

The calculations are carried for the ground ($1s$) and the lowest excited states ($2s$, $2p_{1/2}$, and $2p_{3/2}$). 
In the present work we classify the resonance states by the atomic states with which they are coincident in 
the zero-field limit ($F = 0$). In nonrelativistic studies of the Stark effect, 
another notation, which is based on parabolic quantum numbers $(n_1, n_2, m_L)$~\cite{LL}, 
is usually used. For the states considered there is the following correspondence between the notations:
$1s$, $2s$, $2p_{1/2}$, and $2p_{3/2}$ match $(0, 0, 0)$, $(0, 1, 0)$, 
$(0, 0, 1)$, and $(1, 0, 0)$, respectively. 
\begin{figure}[ht]
    \centering
    \includegraphics{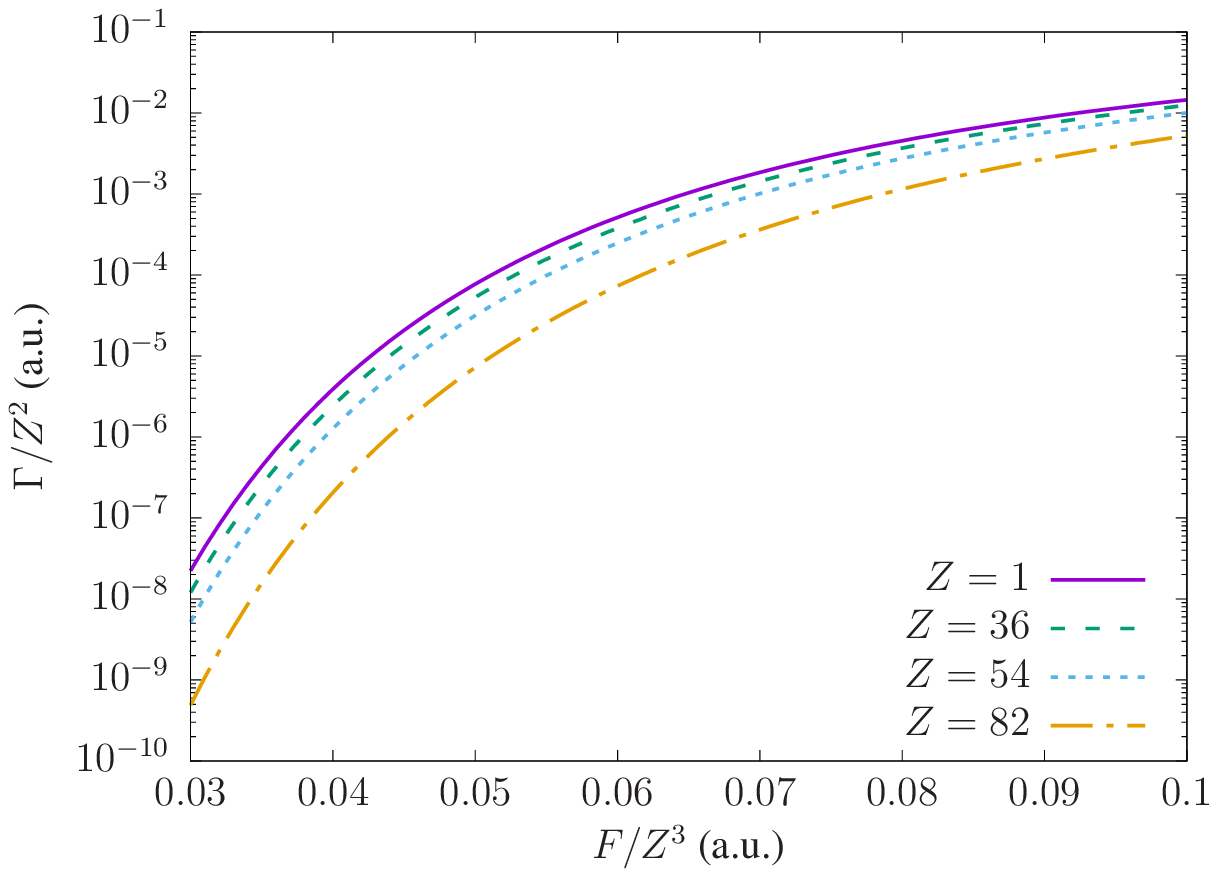}
    \caption{Relativistic Stark width $\Gamma$ of the ground state  
    of the hydrogenlike ion with nuclear charge $Z$ in presence of a 
    uniform electric field $F$. In the nonrelativistic limit, the values for all 
    $Z$ should be the same. The difference is caused by the relativistic effects.}
    \label{fig:widths}
\end{figure}
The results obtained for the Stark shift $\Delta E$ and 
Stark width~$\Gamma$ of the ground state 
and their nonrelativistic counterparts are presented 
in Table~\ref{tab:1s}. The nonrelativistic values for hydrogen are taken from Ref.~\cite{Benassi_80} 
and those for $Z \neq 1$ are derived via scaling of the hydrogen ones. 
As can be seen from the table, the relative difference between the relativistic and 
nonrelativistic Stark shift values is almost the same for all considered field strengths~$F$
and grows with $Z$. 
%There is much more drastic dependence on $F$ of the width. 
All the relativistic width values are smaller than the nonrelativistic ones, 
and the difference is larger for weaker fields and higher $Z$. 
For the lead ion ($Z = 82$) for $F \leqslant 0.04 \times Z^3$ the relativistic width value is 
suppressed relative to the nonrelativistic one by more than one order of magnitude.

\begin{figure}[H]
    \centering
    \includegraphics{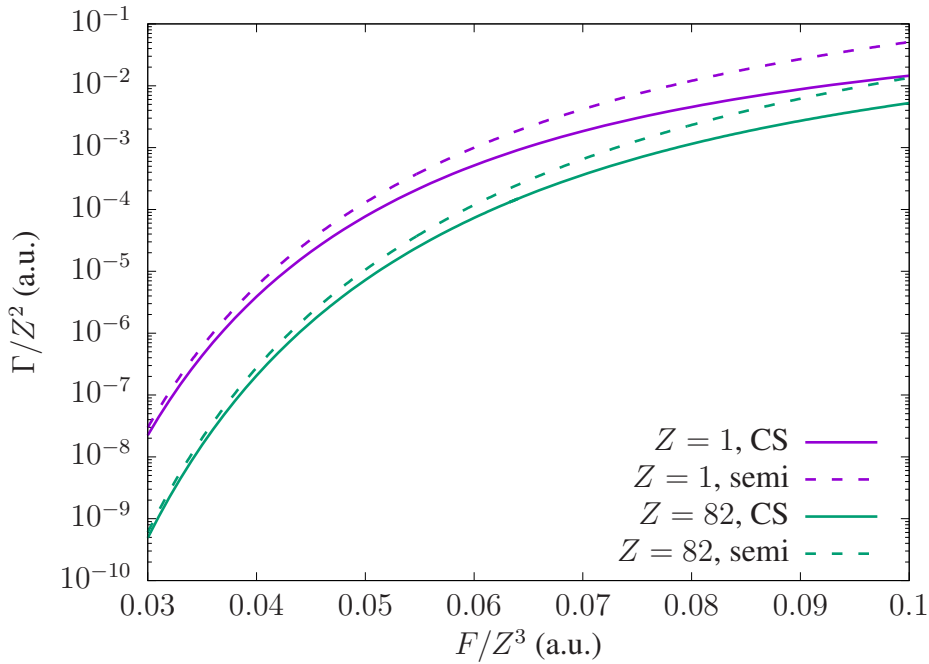}
    \caption{Relativistic Stark width $\Gamma$ for the ground state  
    of the hydrogenlike ion with nuclear charge $Z$ in the presence of a 
    uniform electric field $F$. Solid lines show the results obtained with the 
    complex-scaling method and dashed lines the values calculated using the semiclassical 
    theory. The upper curves correspond to $Z = 1$ and the lower ones correspond to $Z = 82$.}
    \label{fig:cs_vs_semi}
\end{figure}

In order to better illustrate the dependence of the relativistic effects on $Z$ and $F$, 
we present the scaled width $\Gamma / Z^2$ 
as a function of the scaled field strength $F/Z^3$ for several $Z$ in Fig.~\ref{fig:widths}. 
In the nonrelativistic limit for the pointlike nuclei, all the curves should be the same. 
The difference is caused by the relativistic effects. 
As one can see from the figure, the divergence of the curves is larger for the weaker fields
and this behavior becomes more pronounced for higher $Z$.
The fact that the relativistic corrections have more impact for the weaker fields
seems paradoxical. This phenomenon was discovered previously using the semiclassical
approximation~\cite{Milosevic_02, Milosevic_02_2, Popov_04} and found to be a consequence 
of a relativistic increase of the binding energy. 
The CS results obtained for $Z = 1$ and $82$ are compared to the semiclassical ones 
in Fig.~\ref{fig:cs_vs_semi}. The semiclassical values were calculated according to the equation~(36)
from Ref.~\cite{Milosevic_02_2}. As one can see, the semiclassical theory indeed provides the qualitatively 
correct dependence of the width on the field strength. 
However, the semiclassical values are systematically larger than the CS ones and quantitatively valid only 
for small $F$. Such an overestimation of the width by the semiclassical approximation is already known in the 
nonrelativistic case (see, for example, Ref.~\cite{Batishchev_10}). The confirmed relativistic suppression 
of the Stark width means that a heavy ion exposed to the electric field can be much more stable than 
the results obtained from the nonrelativistic calculations.

In Tables~\ref{tab:2p}, \ref{tab:2s}, and \ref{tab:2p_32} we present the results 
for the excited 2$p_{1/2}$, 2$s$, and 2$p_{3/2}$ states respectively. 
As one can see from the tables, for 2$p_{1/2}$ and 2$s$ resonances the 
relativistic widths are also suppressed with respect to the nonrelativistic 
ones and the difference is larger for weaker field~$F$ and higher $Z$. 
The 2$p_{3/2}$ state, however, is a notable exception. For $Z \geqslant 18$
and $F/Z^3 = 0.05$~a.u. the relativistic width of 2$p_{3/2}$ state is larger than the nonrelativistic
value and for $Z = 82$ the difference is more than one order of magnitude. 
A possible explanation for such a drastic discrepancy is the influence of the spin-orbital 
interaction, which can play a significant role for small $F$. 
This suggests that this effect can be found using the two-component calculation 
methods. In the case of heavy ions, however, their accuracy is quite limited. 

\begin{figure}[H]
    \centering
    \includegraphics{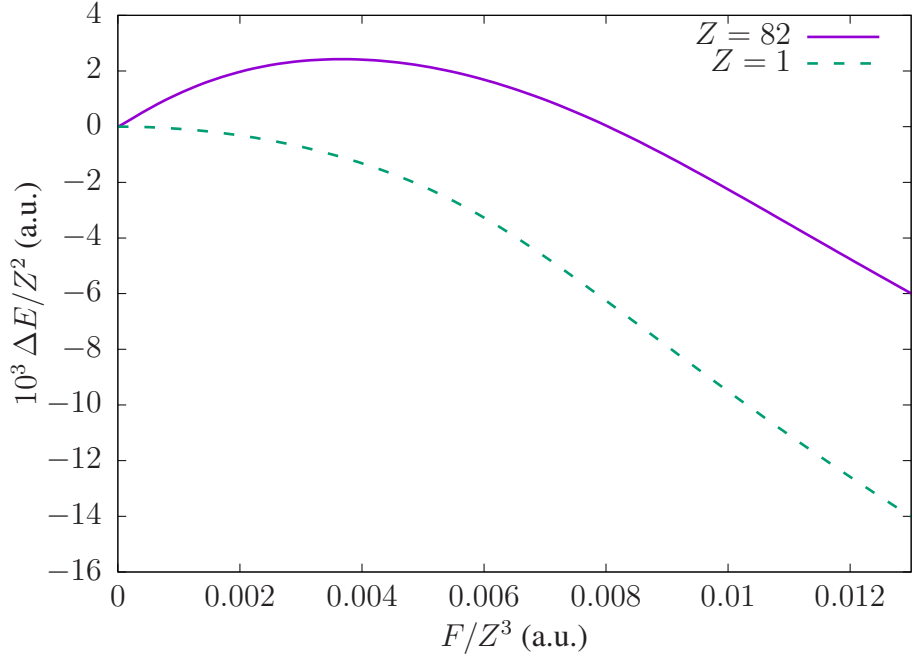}
    \caption{Stark shift $\Delta E$ for the 2$s$ state of the hydrogenlike ion 
    with the nuclear charge $Z$ in the presence of a uniform electric field~$F$.}
    \label{fig:pos_2s}
\end{figure}

Almost all the presented relativistic values of the energy shift are slightly smaller than the 
nonrelativistic counterparts. There is a more complicated situation for the $2s$ state.
As can be seen from Table~\ref{tab:2s}, for $Z = 82$ the relativistic and 
nonrelativistic values have the opposite signs. The comparison between results for $Z = 82$ scaled by $1/Z^2$
and the corresponding values for $Z = 1$ is shown in Fig.~\ref{fig:pos_2s}. The difference in behavior 
is explained by the relativistic effects since for $Z = 1$ they are almost negligible. 
It should be noted that the opposite sign for the relativistic value of the Stark shift
was previously reported in Ref.~\cite{Rozenbaum_14} for an argon ion.
\begin{center}
\begin{longtable}{
  S[table-format=1.2]
  S[table-format=3.9(1)e2, table-align-exponent = false]
  S[table-format=3.7(1)e2, table-align-exponent = false]
  S[table-format=3.7(1)e2, table-align-exponent = false]
  S[table-format=3.7(1)e2, table-align-exponent = false]}\\
\caption{Stark shifts~$\Delta E$ and widths~$\Gamma$ of the ground-state resonance of hydrogenlike ions with 
the nuclear charge~$Z$ as functions of the field strength~$F$. The relativistic results were obtained in this work.
The nonrelativistic results are the hydrogen values from Ref.~\cite{Benassi_80} multiplied by $Z^2$.}
 \label{tab:1s} \\ \hline\hline
            & \multicolumn{2}{c}{Relativistic} &             \multicolumn{2}{c}{Non-relativistic~~~~~}\\
            {$F / Z^3$~(a.u.)} & {$\Delta E$ (a.u.)} & {$-\Gamma/2$ (a.u.)}            & {$\Delta E$ (a.u.)}  & {$-\Gamma/2$ (a.u.)}\\ 
\hline
\multicolumn{5}{c}{ {$Z$ = 1, $R_{\rm RMS}$ = 0.8775~fm} }\\
   0.03 & -2.0741555(1)e-03 & -1.11825(2)e-08  &-2.074273e-03 &    -1.11880e-08 \\
   0.04 & -3.7713727(1)e-03 & -1.945666(6)e-06  &-3.771591e-03 &    -1.94635e-06 \\
   0.05 & -6.1050580(1)e-03 & -3.858129(5)e-05  &-6.105425e-03 &   -3.859208e-05 \\
   0.06 & -9.2028811(1)e-03 & -2.574804(1)e-04  &-9.203451e-03 &  -2.5753874e-04 \\
   0.07 & -1.30759641(1)e-02 & -9.235128(2)e-04  &-1.307677e-02 &  -9.2368428e-04 \\
   0.08 & -1.75595902(1)e-02 & -2.2694781(2)e-03  &-1.756062e-02 &  -2.2698288e-03 \\
   0.09 & -2.24115841(2)e-02 & -4.3914109(4)e-03  &-2.241281e-02 &  -4.3919872e-03 \\
    0.1 & -2.74167896(5)e-02 & -7.2682294(4)e-03  &-2.741818e-02 &  -7.2690568e-03 \\

\multicolumn{5}{c}{ {$Z$ = 10, $R_{\rm RMS}$ = 3.0055~fm} }\\
   0.03 & -2.0625762(1)e-01 & -1.06837(2)e-06  &-2.074273e-01 &    -1.11880e-06 \\
   0.04 & -3.7497908(1)e-01 & -1.879052(6)e-04  &-3.771591e-01 &    -1.94635e-04 \\
   0.05 & -6.0687382(1)e-01 & -3.752554(5)e-03  &-6.105425e-01 &   -3.859208e-03 \\
   0.06 & -9.1465468(1)e-01 & -2.517546(1)e-02  &-9.203451e-01 &  -2.5753874e-02 \\
   0.07 & -1.29966685(1)e+00 & -9.066438(2)e-02  &-1.307677e+00 &  -9.2368428e-02 \\
   0.08 & -1.74579789(1)e+00 & -2.2349116(2)e-01  &-1.756062e+00 &  -2.2698288e-01 \\
   0.09 & -2.22904989(2)e+00 & -4.3345338(3)e-01  &-2.241281e+00 &  -4.3919872e-01 \\
    0.1 & -2.72795663(5)e+00 & -7.1865027(4)e-01  &-2.741818e+00 &  -7.2690568e-01 \\

\multicolumn{5}{c}{ {$Z$ = 18, $R_{\rm RMS}$ = 3.4028~fm} }\\
   0.03 & -6.598073(1)e-01 & -3.11911(7)e-06  &-6.720643e-01 &    -3.62491e-06 \\
   0.04 & -1.1991590(1)e+00 & -5.62253(2)e-04  &-1.221995e+00 &    -6.30617e-04 \\
   0.05 & -1.9397447(1)e+00 & -1.141133(2)e-02  &-1.978158e+00 &   -1.250383e-02 \\
   0.06 & -2.9223297(1)e+00 & -7.747931(4)e-02  &-2.981918e+00 &  -8.3442552e-02 \\
   0.07 & -4.1529243(1)e+00 & -2.8161645(6)e-01  &-4.236872e+00 &  -2.9927371e-01 \\
   0.08 & -5.5819469(1)e+00 & -6.9910647(8)e-01  &-5.689640e+00 &  -7.3542452e-01 \\
   0.09 & -7.1332880(1)e+00 & -1.3630835(1)e+00  &-7.261750e+00 &  -1.4230039e+00 \\
    0.1 & -8.7377925(2)e+00 & -2.2689107(1)e+00  &-8.883489e+00 &  -2.3551744e+00 \\

\multicolumn{5}{c}{ {$Z$ = 36, $R_{\rm RMS}$ = 4.1835~fm} }\\
   0.03 & -2.4936111(1)e+00 &  -7.8166(3)e-06  &-2.688257e+00 &    -1.44996e-05 \\
   0.04 & -4.5259171(1)e+00 & -1.572824(6)e-03  &-4.887982e+00 &    -2.52247e-03 \\
   0.05 & -7.3048551(1)e+00 & -3.430401(5)e-02  &-7.912631e+00 &   -5.001534e-02 \\
   0.06 & -1.09845515(1)e+01 & -2.456480(2)e-01  &-1.192767e+01 &  -3.3377021e-01 \\
   0.07 & -1.56140755(1)e+01 & -9.304429(2)e-01  &-1.694749e+01 &  -1.1970948e+00 \\
   0.08 & -2.10396505(1)e+01 & -2.3841355(3)e+00  &-2.275856e+01 &  -2.9416981e+00 \\
   0.09 & -2.69874032(2)e+01 & -4.7610474(4)e+00  &-2.904700e+01 &  -5.6920155e+00 \\
    0.1 & -3.31897931(5)e+01 & -8.0689585(6)e+00  &-3.553395e+01 &  -9.4206976e+00 \\

\multicolumn{5}{c}{ {$Z$ = 54, $R_{\rm RMS}$ = 4.7964~fm} }\\
   0.03 & -5.075628(1)e+00 &  -7.5824(6)e-06  &-6.048579e+00 &    -3.26242e-05 \\
   0.04 & -9.192809(1)e+00 & -1.856289(8)e-03  &-1.099796e+01 &    -5.67556e-03 \\
   0.05 & -1.4783894(1)e+01 & -4.599133(8)e-02  &-1.780342e+01 &   -1.125345e-01 \\
   0.06 & -2.2152360(1)e+01 & -3.617239(3)e-01  &-2.683726e+01 &  -7.5098297e-01 \\
   0.07 & -3.1473089(1)e+01 & -1.4738926(5)e+00  &-3.813185e+01 &  -2.6934634e+00 \\
   0.08 & -4.2554122(1)e+01 & -3.9978229(6)e+00  &-5.120676e+01 &  -6.6188207e+00 \\
   0.09 & -5.4907367(1)e+01 & -8.3398840(8)e+00  &-6.535575e+01 &  -1.2807035e+01 \\
    0.1 & -6.7983841(1)e+01 & -1.4612320(1)e+01  &-7.995140e+01 &  -2.1196570e+01 \\

\multicolumn{5}{c}{ {$Z$ = 82, $R_{\rm RMS}$ = 5.5012~fm} }\\
   0.03 & -8.935694(1)e+00 &   -1.646(1)e-06  &-1.394741e+01 &    -7.52281e-05 \\
   0.04 & -1.6112143(1)e+01 & -6.92395(5)e-04  &-2.536018e+01 &    -1.30873e-02 \\
   0.05 & -2.5712020(1)e+01 & -2.427590(6)e-02  &-4.105288e+01 &   -2.594931e-01 \\
   0.06 & -3.8160498(1)e+01 & -2.455934(3)e-01  &-6.188400e+01 &  -1.7316905e+00 \\
   0.07 & -5.3920546(1)e+01 & -1.2177878(7)e+00  &-8.792817e+01 &  -6.2108531e+00 \\
   0.08 & -7.3126730(1)e+01 & -3.863345(1)e+00  &-1.180776e+02 &  -1.5262329e+01 \\
   0.09 & -9.5381168(1)e+01 & -9.125540(1)e+00  &-1.507037e+02 &  -2.9531722e+01 \\
    0.1 & -1.19906312(1)e+02 & -1.7627242(2)e+01  &-1.843598e+02 &  -4.8877138e+01 \\

\hline\hline

\end{longtable}
\end{center}
\begin{center}
\begin{longtable}{
  S[table-format=1.2]
  S[table-format=3.9(1)e2, table-align-exponent = false]
  S[table-format=3.7(1)e2, table-align-exponent = false]
  S[table-format=3.7(1)e2, table-align-exponent = false]
  S[table-format=3.7(1)e2, table-align-exponent = false]}\\
\caption{Stark shifts~$\Delta E$ and widths~$\Gamma$ of the $2p_{1/2}$ resonance of hydrogenlike ions with 
the nuclear charge~$Z$ as functions of the field strength~$F$. The relativistic results were obtained in this work.
The nonrelativistic results are the hydrogen values from Ref.~\cite{Rao_94} multiplied by $Z^2$.}
 \label{tab:2p} \\ \hline\hline
            & \multicolumn{2}{c}{Relativistic} &             \multicolumn{2}{c}{Non-relativistic~~~~~}\\
            {$F / Z^3$~(a.u.)} & {$\Delta E$ (a.u.)} & {$-\Gamma/2$ (a.u.)}            & {$\Delta E$ (a.u.)}  & {$-\Gamma/2$ (a.u.)}\\ 
\hline
\multicolumn{5}{c}{ {$Z$ = 1, $R_{\rm RMS}$ = 0.8775~fm} }\\
  0.005 & -1.76176897(8)e-02 & -5.29586(3)e-05  &-1.761861e-02 &  -5.2972236e-05 \\
   0.01 & -4.10926693(5)e-02 & -5.442161(2)e-03  &-4.109400e-02 &  -5.4425560e-03 \\
   0.02 & -8.1680516(7)e-02 & -3.0391832(5)e-02  &-8.182220e-02 &  -3.0392847e-02 \\
   0.03 & -1.1514503(2)e-01 & -5.981855(6)e-02  &-1.151471e-01 &   -5.982000e-02 \\

\multicolumn{5}{c}{ {$Z$ = 10, $R_{\rm RMS}$ = 3.0055~fm} }\\
  0.005 & -1.75268320(8)e+00 & -5.16194(3)e-03  &-1.761861e+00 &  -5.2972236e-03 \\
   0.01 & -4.09605836(5)e+00 & -5.403122(2)e-01  &-4.109400e+00 &  -5.4425560e-01 \\
   0.02 & -8.1509830(7)e+00 & -3.0290352(5)e+00  &-8.182220e+00 &  -3.0392847e+00 \\
   0.03 & -1.1493599(2)e+01 & -5.966553(6)e+00  &-1.151471e+01 &   -5.982000e+00 \\

\multicolumn{5}{c}{ {$Z$ = 18, $R_{\rm RMS}$ = 3.4028~fm} }\\
  0.005 & -5.6121159(3)e+00 & -1.576507(8)e-02  &-5.708429e+00 &  -1.7163004e-02 \\
   0.01 & -1.31737804(2)e+01 & -1.7218961(6)e+00  &-1.331446e+01 &  -1.7633881e+00 \\
   0.02 & -2.6283190(2)e+01 & -9.739293(2)e+00  &-2.651039e+01 &  -9.8472823e+00 \\
   0.03 & -3.7085071(6)e+01 & -1.921889(2)e+01  &-3.730767e+01 &   -1.938168e+01 \\

\multicolumn{5}{c}{ {$Z$ = 36, $R_{\rm RMS}$ = 4.1835~fm} }\\
  0.005 & -2.1296520(1)e+01 & -4.78561(3)e-02  &-2.283372e+01 &  -6.8652017e-02 \\
   0.01 & -5.09632086(5)e+01 & -6.382976(2)e+00  &-5.325783e+01 &  -7.0535526e+00 \\
   0.02 & -1.02887904(7)e+02 & -3.7633130(6)e+01  &-1.060416e+02 &  -3.9389129e+01 \\
   0.03 & -1.4560076(2)e+02 & -7.488076(7)e+01  &-1.492307e+02 &   -7.752672e+01 \\

\multicolumn{5}{c}{ {$Z$ = 54, $R_{\rm RMS}$ = 4.7964~fm} }\\
  0.005 & -4.3657867(2)e+01 & -6.26878(4)e-02  &-5.137586e+01 &  -1.5446704e-01 \\
   0.01 & -1.078292818(7)e+02 & -1.2420509(5)e+01  &-1.198301e+02 &  -1.5870493e+01 \\
   0.02 & -2.2256483(1)e+02 & -7.947587(1)e+01  &-2.385935e+02 &  -8.8625541e+01 \\
   0.03 & -3.1676073(3)e+02 & -1.606622(1)e+02  &-3.357690e+02 &   -1.744351e+02 \\

\multicolumn{5}{c}{ {$Z$ = 82, $R_{\rm RMS}$ = 5.5012~fm} }\\
  0.005 & -7.8941197(3)e+01 & -2.35907(2)e-02  &-1.184675e+02 &  -3.5618531e-01 \\
   0.01 & -2.07353215(1)e+02 & -1.773695(1)e+01  &-2.763161e+02 &  -3.6595747e+01 \\
   0.02 & -4.5752851(1)e+02 & -1.5173717(2)e+02  &-5.501725e+02 &  -2.0436150e+02 \\
   0.03 & -6.6341661(1)e+02 & -3.231929(2)e+02  &-7.742493e+02 &   -4.022297e+02 \\

\hline\hline

\end{longtable}
\end{center}
\begin{center}
\begin{longtable}{
  S[table-format=1.2]
  S[table-format=3.9(1)e2, table-align-exponent = false]
  S[table-format=3.7(1)e2, table-align-exponent = false]
  S[table-format=3.7(1)e2, table-align-exponent = false]
  S[table-format=3.7(1)e2, table-align-exponent = false]}\\
\caption{Stark shifts~$\Delta E$ and widths~$\Gamma$ of the $2s$ resonance of the hydrogenlike ions with 
the nuclear charge~$Z$ as functions of the field strength~$F$. The relativistic results were obtained in this work.
The nonrelativistic results are the hydrogen values from Ref.~\cite{Rao_94} multiplied by $Z^2$.}
 \label{tab:2s} \\ \hline\hline
            & \multicolumn{2}{c}{Relativistic} &             \multicolumn{2}{c}{Non-relativistic~~~~~}\\
            {$F / Z^3$~(a.u.)} & {$\Delta E$ (a.u.)} & {$-\Gamma/2$ (a.u.)}            & {$\Delta E$ (a.u.)}  & {$-\Gamma/2$ (a.u.)}\\ 
\hline
\multicolumn{5}{c}{ {$Z$ = 1, $R_{\rm RMS}$ = 0.8775~fm} }\\
  0.005 & -2.145930(1)e-03 & -1.30722(3)e-05  &-2.146613e-03 &  -1.3076437e-05 \\
   0.01 & -9.5239060(7)e-03 & -3.138327(6)e-03  &-9.524887e-03 &  -3.1386570e-03 \\
   0.02 & -2.172828(4)e-02 & -2.185255(6)e-02  &-2.172946e-02 &  -2.1853572e-02 \\
   0.03 & -2.83555(4)e-02 & -4.44219(2)e-02  &-2.835714e-02 &  -4.4424040e-02 \\

\multicolumn{5}{c}{ {$Z$ = 10, $R_{\rm RMS}$ = 3.0055~fm} }\\
  0.005 & -2.078337(1)e-01 & -1.26770(3)e-03  &-2.146613e-01 &  -1.3076437e-03 \\
   0.01 & -9.4265659(7)e-01 & -3.106086(6)e-01  &-9.524887e-01 &  -3.1386570e-01 \\
   0.02 & -2.161869(4)e+00 & -2.174493(6)e+00  &-2.172946e+00 &  -2.1853572e+00 \\
   0.03 & -2.82390(4)e+00 & -4.42479(2)e+00  &-2.835714e+00 &  -4.4424040e+00 \\

\multicolumn{5}{c}{ {$Z$ = 18, $R_{\rm RMS}$ = 3.4028~fm} }\\
  0.005 & -6.236963(4)e-01 & -3.85273(9)e-03  &-6.955025e-01 &  -4.2367655e-03 \\
   0.01 & -2.9824595(2)e+00 & -9.82897(2)e-01  &-3.086063e+00 &  -1.0169249e+00 \\
   0.02 & -6.92365(1)e+00 & -6.96632(2)e+00  &-7.040345e+00 &  -7.0805573e+00 \\
   0.03 &  -9.0636(1)e+00 & -1.420845(7)e+01  &-9.187712e+00 &  -1.4393389e+01 \\

\multicolumn{5}{c}{ {$Z$ = 36, $R_{\rm RMS}$ = 4.1835~fm} }\\
  0.005 & -1.625689(2)e+00 & -1.27550(3)e-02  &-2.782010e+00 &  -1.6947062e-02 \\
   0.01 & -1.06592307(2)e+01 & -3.534916(7)e+00  &-1.234425e+01 &  -4.0676994e+00 \\
   0.02 & -2.625980(4)e+01 & -2.648461(7)e+01  &-2.816138e+01 &  -2.8322229e+01 \\
   0.03 & -3.47389(4)e+01 & -5.45927(2)e+01  &-3.675085e+01 &  -5.7573556e+01 \\

\multicolumn{5}{c}{ {$Z$ = 54, $R_{\rm RMS}$ = 4.7964~fm} }\\
  0.005 & -3.94071(3)e-01 & -2.53892(4)e-02  &-6.259523e+00 &  -3.8130889e-02 \\
   0.01 & -1.9015147(1)e+01 & -6.56114(1)e+00  &-2.777457e+01 &  -9.1523237e+00 \\
   0.02 & -5.341589(5)e+01 & -5.43389(1)e+01  &-6.336311e+01 &  -6.3725015e+01 \\
   0.03 & -7.22343(8)e+01 & -1.142448(4)e+02  &-8.268941e+01 &  -1.2954050e+02 \\

\multicolumn{5}{c}{ {$Z$ = 82, $R_{\rm RMS}$ = 5.5012~fm} }\\
  0.005 & 1.4704343(5)e+01 & -2.44920(2)e-02  &-1.443382e+01 &  -8.7925960e-02 \\
   0.01 & -1.5160798(8)e+01 & -8.80437(2)e+00  &-6.404534e+01 &  -2.1104329e+01 \\
   0.02 & -8.8440982(6)e+01 & -9.59044(1)e+01  &-1.461089e+02 &  -1.4694342e+02 \\
   0.03 & -1.30752(1)e+02 & -2.144554(3)e+02  &-1.906734e+02 &  -2.9870724e+02 \\

\hline\hline

\end{longtable}
\end{center}
% %
\begin{center}
\begin{longtable}{
  S[table-format=1.2]
  S[table-format=3.9(1)e2, table-align-exponent = false]
  S[table-format=3.7(1)e2, table-align-exponent = false]
  S[table-format=3.7(1)e2, table-align-exponent = false]
  S[table-format=3.7(1)e2, table-align-exponent = false]}\\
\caption{Stark shifts~$\Delta E$ and widths~$\Gamma$ of the $2p_{3/2}$ resonance of hydrogenlike ions with 
the nuclear charge~$Z$ as functions of the field strength~$F$. The relativistic results were obtained in this work. The nonrelativistic results are the hydrogen values from Ref.~\cite{Rao_94} multiplied by $Z^2$.}
 \label{tab:2p_32} \\ \hline\hline
            & \multicolumn{2}{c}{Relativistic} &             \multicolumn{2}{c}{Non-relativistic~~~~~}\\
            {$F / Z^3$~(a.u.)} & {$\Delta E$ (a.u.)} & {$-\Gamma/2$ (a.u.)}            & {$\Delta E$ (a.u.)}  & {$-\Gamma/2$ (a.u.)}\\ 
\hline
\multicolumn{5}{c}{ {$Z$ = 1, $R_{\rm RMS}$ = 0.8775~fm} }\\
  0.005 & 1.2936853(1)e-02 &  -2.8636(2)e-06  &1.293808e-02 &   -2.864697e-06 \\
   0.01 & 2.1104324(3)e-02 & -1.639384(7)e-03  &2.110544e-02 &  -1.6396395e-03 \\
   0.02 &  3.60149(4)e-02 & -1.54447(3)e-02  &3.601573e-02 &  -1.5446311e-02 \\
   0.03 &   5.4275(3)e-02 &  -3.3259(2)e-02  &5.428097e-02 &   -3.326218e-02 \\

\multicolumn{5}{c}{ {$Z$ = 10, $R_{\rm RMS}$ = 3.0055~fm} }\\
  0.005 & 1.2815660(1)e+00 &  -2.7991(2)e-04  &1.293808e+00 &   -2.864697e-04 \\
   0.01 & 2.0994469(3)e+00 & -1.615696(7)e-01  &2.110544e+00 &  -1.6396395e-01 \\
   0.02 &  3.58815(4)e+00 & -1.53364(3)e+00  &3.601573e+00 &  -1.5446311e+00 \\
   0.03 &   5.4099(3)e+00 &  -3.3070(2)e+00  &5.428097e+00 &   -3.326218e+00 \\

\multicolumn{5}{c}{ {$Z$ = 18, $R_{\rm RMS}$ = 3.4028~fm} }\\
  0.005 & 4.0630599(4)e+00 &  -9.5876(8)e-04  &4.191937e+00 &   -9.281618e-04 \\
   0.01 & 6.7209719(7)e+00 & -5.06648(2)e-01  &6.838164e+00 &  -5.3124321e-01 \\
   0.02 &  1.15271(1)e+01 &  -4.8903(1)e+00  &1.166910e+01 &  -5.0046046e+00 \\
   0.03 &   1.7398(1)e+01 & -1.05775(6)e+01  &1.758704e+01 &   -1.077695e+01 \\

\multicolumn{5}{c}{ {$Z$ = 36, $R_{\rm RMS}$ = 4.1835~fm} }\\
  0.005 & 1.4683176(1)e+01 & -1.09265(4)e-02  &1.676775e+01 &   -3.712647e-03 \\
   0.01 & 2.54275429(5)e+01 & -1.769986(7)e+00  &2.735265e+01 &  -2.1249728e+00 \\
   0.02 &  4.43478(5)e+01 & -1.82393(3)e+01  &4.667638e+01 &  -2.0018418e+01 \\
   0.03 &   6.7277(3)e+01 &  -3.9986(2)e+01  &7.034814e+01 &   -4.310779e+01 \\

\multicolumn{5}{c}{ {$Z$ = 54, $R_{\rm RMS}$ = 4.7964~fm} }\\
  0.005 & 2.7057912(3)e+01 & -9.53525(9)e-02  &3.772743e+01 &   -8.353456e-03 \\
   0.01 & 5.1356073(9)e+01 & -3.343263(8)e+00  &6.154347e+01 &  -4.7811889e+00 \\
   0.02 &  9.27424(9)e+01 & -3.64837(3)e+01  &1.050219e+02 &  -4.5041442e+01 \\
   0.03 &  1.42198(4)e+02 &  -8.1759(4)e+01  &1.582833e+02 &   -9.699252e+01 \\

\multicolumn{5}{c}{ {$Z$ = 82, $R_{\rm RMS}$ = 5.5012~fm} }\\
  0.005 & 3.1618260(7)e+01 & -7.69521(3)e-01  &8.699562e+01 &   -1.926222e-02 \\
   0.01 & 8.194625(2)e+01 & -8.91994(2)e+00  &1.419130e+02 &  -1.1024936e+01 \\
   0.02 &  1.69924(1)e+02 & -6.55436(4)e+01  &2.421697e+02 &  -1.0386099e+02 \\
   0.03 &  2.71469(1)e+02 & -1.51054(7)e+02  &3.649853e+02 &   -2.236549e+02 \\

\hline\hline

\end{longtable}
\end{center}
\section{CONCLUSION}
In the present work, we calculated the relativistic positions and widths of the Stark resonances
in hydrogenlike ions using the complex-scaling method.
The calculations were performed for the $1s$, $2s$, $2p_{1/2}$, and $2p_{3/2}$ states 
of several ions between $Z = 1$ and $Z = 82$. The obtained results show the importance 
of relativistic effects.
The comparison between the relativistic and nonrelativistic
values leads to the conclusion that the nonrelativistic calculations are unreliable for heavy
ions. The difference is especially drastic for the Stark widths and can be more than one order of magnitude. 
It is also worth noting that the influence of the relativistic effects is larger for smaller values of the 
external electric field, which are easier to achieve experimentally. 

The performed calculations have confirmed the relativistic suppression of the ground state width 
which was previously shown in Refs.~\cite{Milosevic_02, Milosevic_02_2} using the semiclassical
method. In the present work, the existence of the same effect was demonstrated for $2s$ and $2p_{1/2}$ states.
However, the situation may be the opposite for the $2p_{3/2}$ state for sufficiently high $Z$ and weak 
external field. This emphasizes the importance of relativistic consideration of the Stark effect in heavy ions.
It should be noted that despite the fact that the semiclassical theory can provide a qualitatively
correct description of the relativistic effects on the energy width, its quantitative predictions can 
be quite far from the exact values. 

Our consideration was restricted to hydrogenlike ions in the inertial reference frame exposed
to a uniform constant electrical field. Real experimental conditions can be much more complicated and 
include a magnetic field, ion acceleration, and other factors. In order to estimate the influence of 
all these factors further development is required. 
Nevertheless, we expect that the obtained results will be useful for 
future experiments with heavy partially stripped ions in strong electric fields.

\section{ACKNOWLEDGMENTS}
This work was supported by the
Foundation for the Advancement of Theoretical Physics and Mathematics “BASIS”.
%
% END ACKNOWLEDGEMENTS =================================================
%
%=====================================================================
%

% ======================================================================
%

\begin{thebibliography}{99}
\bibitem{Traubenberg_81} H.~Rausch v. Traubenberg, R.~Gebauer, and G.~Lewin, Naturwissenschaften {\bf 18}, 417 (1930).
%
\bibitem{Stebbings_76} R.~F.~Stebbings, Science {\bf 193}, 537 (1976).
%
\bibitem{Littman_76} M.~G.~Littman, M.~L.~Zimmerman, and D.~Kleppner, Phys. Rev.
Lett. {\bf 37}, 486 (1976).
%
\bibitem{Koch_81} P.~M.~Koch and D.~R.~Mariani,
Phys. Rev. Lett. {\bf 46}, 1275 (1981).
%
\bibitem{Bergman_84} T.~Bergeman, C.~Harvey, K.~B.~Butterfield, H.~C.~Bryant, 
D.~A.~Clark, P.~A.~M.~Gram, D.~MacArthur, M.~Davis, J.~B.~Donahue, 
J.~Dayton, and W.~W.~Smith, Phys.~Rev.~Lett. {\bf 53}, 775 (1984).
%
\bibitem{Stodolna_13} A.~S.~Stodolna, A.~Rouz\'ee, F.~L\'epine, S.~Cohen, F.~Robicheaux,
A.~Gijsbertsen, J.~H.~Jungmann, C.~Bordas, and M.~J.~J.~Vrakking,
Phys. Rev. Lett. {\bf 110}, 213001 (2013).
%
\bibitem{Hehenberger_74} M.~Hehenberger, H.~V.~McIntosh, and E.~Br\"andas, 
Phys. Rev A {\bf 10}, 1494 (1974).
%
\bibitem{Zapryagaev_78} S.~A.~Zapryagaev, Opt. Spectrosk. {\bf 44}, 892 (1978) [in Russian].
%
\bibitem{Benassi_80} L.~Benassi and V.~Grecchi, J. Phys. B {\bf 13}, 911 (1980).
%
\bibitem{Farreley_83} D.~Farrelly and W.~P.~Reinhardt, J.~Phys.~B {\bf 16}, 2103 (1983).
%
\bibitem{Gallas_82} J.~A.~C.~Gallas, H.~Walther, and E.~Werner, Phys. Rev. A {\bf 26}, 1775 (1982).
%
\bibitem{Damburg_76} R.~J.~Damburg and V.~V.~Kolosov, J. Phys. B {\bf 9}, 3149 (1976).
%
\bibitem{Damburg_78} R.~J.~Damburg and V.~V.~Kolosov, J. Phys. B {\bf 11}, 1921 (1978).
%
\bibitem{Kolosov_87} V.~V.~Kolosov, J.~Phys. B {\bf 20}, 2359 (1987).
%
\bibitem[]{Lai_81} C.~S.~Lai, Phys. Lett. {\bf 83A}, 322 (1981).
%
\bibitem[]{Kolosov_83} V. V. Kolosov, J. Phys. B {\bf 16}, 25 (1983).
%
\bibitem{Maquet_83} A. Maquet, S.-I. Chu, and W. P. Reinhardt, Phys. Rev. A {\bf 27}, 2946 (1983).
%
\bibitem{Lin_11} C. Y. Lin and Y. K. Ho, J. Phys. B {\bf 44}, 175001 (2011).
%
\bibitem{Rao_94} J.~Rao, W.~Liu, and B.~Li, Phys. Rev. A {\bf 50}, 1916 (1994). 
%
\bibitem{Fernandez_96} F.~M.~Fernández, Phys. Rev. A {\bf 54}, 1206 (1996).
%
\bibitem{Jenschura_01} U.~D. Jentschura, Phys. Rev. A {\bf 64}, 013403 (2001).
%Andreev,
\bibitem{Ivanov_04} I.~A.~Ivanov and Y.~K.~Ho,  Phys. Rev A {\bf 69}, 023407 (2004).
%==============================quasiclassical================================================%
\bibitem{Milosevic_02} N.~Milosevic, V.~P.~Krainov, and T.~Brabec,
Phys. Rev. Lett. {\bf 89}, 193001 (2002).
%DOI: 10.1103/PhysRevLett.89.193001
%
\bibitem{Milosevic_02_2} N.~Milosevic, V.~P.~Krainov, and T.~Brabec,
J. Phys. B {\bf 35}, 3515 (2002). 
%DOI: 10.1088/0953-4075/35/16/311
%
\bibitem{Popov_04}
V.~S.~Popov, B.~M.~Karnakov, and V.~D.~Mur, JETP Lett. {\bf 79}, 262 (2004).
%============================================================================================%
\bibitem{Batishchev_10} P.~A.~Batishchev, O.~I.~Tolstikhin, and T.~Morishita,
Phys. Rev. A {\bf 82}, 023416 (2010).
%
\bibitem{Ferandez-Menchero_13}  L. Fern\'andez-Menchero and H.~P.~Summers, 
Phys Rev A {\bf 88}, 022509 (2013).
%
\bibitem{Rozenbaum_14} E.~B.~Rozenbaum, D.~A.~Glazov, V.~M.~Shabaev, 
 K.~E.~Sosnova, and D.~A.~Telnov, Phys. Rev. A {\bf 89}, 012514 (2014).
%
\bibitem{Fernandez_18} F.~M.~Fernández, App. Math. and Comp. {\bf 317}, 101 (2018).
%
\bibitem{Maltsev_21} I.~A.~Maltsev, D.~A.~Tumakov, R.~V.~Popov, and V.~M.~Shabaev, 
Opt. Spectrosk. {\bf 130}, 585 (2022).
%============================================================================================%
\bibitem{Andreev_18} V.~Andreev et al., Nature {\bf 562}, 355 (2018).
\bibitem{Blanchard_23} J.~W.~Blanchard, D.~Budker, D.~DeMille, M.~G.~Kozlov, and L.~V.~Skripnikov,
Phys. Rev. Research {\bf 5}, 013191 (2023).
%
\bibitem{Budker_20} D.~Budker, J.~R.~Crespo L\'opez-Urrutia, A.~Derevianko, V.~V. Flambaum, 
M.~W.~Krasny, A.~Petrenko, S.~Pustelny, A.~Surzhykov, 
V.~A.~Yerokhin, and M.~Zolotorev, Ann. Phys. (Berlin) {\bf 532}, 2000204 (2020).
%===================many-electron autoionization states ================================%
%
\bibitem{Zaytsev_19} V.~A.~Zaytsev, I.~A.~Maltsev, I.~I.~Tupitsyn, 
and V.M.~Shabaev,
Phys. Rev. A {\bf 100}, 052504 (2019).
% DOI: 10.1103/PhysRevA.100.052504
%
\bibitem{Kieslich_04} S.~Kieslich, S.~Schippers, W.~Shi, A.~M\"uller, 
G.~Gwinner, M.~Schnell, A.~Wolf, E.~Lindroth, and M.~Tokman,
Phys. Rev. A {\bf 70}, 0042714 (2004).
% DOI: 10.1103/PhysRevA.70.042714
%
\bibitem{Mueller_18} A.~M\"uller, E.~Lindroth, S.~Bari, A.~Borovik Jr.,
P.-M.~Hillenbrand, K.~Holste, P.~Indelicato, A.~L.~D.~Kilcoyne,
S.~Klumpp, M.~Martins, J.~Viefhaus, P.~Wilhelm, and S.~Schippers,
Phys. Rev. A {\bf 98}, 033416 (2018).
% DOI: 10.1103/PhysRevA.98.033416
%
\bibitem{Zaytsev_20} V.~A.~Zaytsev, I.~A.~Maltsev, I.~I~Tupitsyn, V.~M.~Shabaev, 
and V.~Y.~Ivanov,
Opt. and Spectr. {\bf 128}, 307 (2020). 
% DOI: 10.1134/S0030400X20030200]. 
%==================================== supercritical ===================================%
 \bibitem{Ackad_07_1} E.~Ackad and M.~Horbatsch, 
 Phys. Rev. A {\bf 75}, 022508 (2007).
% DOI: 10.1103/PhysRevA.75.022508
%
 \bibitem{Ackad_07_2} E.~Ackad and M.~Horbatsch,
 Phys. Rev. A {\bf 76}, 022503 (2007).
% DOI: 10.1103/PhysRevA.76.022503
% 
\bibitem{Marsman_11} A.~Marsman and M.~Horbatsch, Phys. Rev. A {\bf 84}, 032517 (2011).
% DOI: 10.1103/PhysRevA.84.032517
%
\bibitem{Maltsev_20}
I.~A.~Maltsev, V.~M.~Shabaev, V.~A.~Zaytsev, 
R.~V.~Popov, and D.~A.~Tumakov,
Opt. Spectrosk. {\bf 128}, 1100 (2020).
%=======================================Q-chem=========================================%
\bibitem{Jagau_18} T.-C.~Jagau, J. Chem. Phys. {\bf 148}, 204102 (2018).
\bibitem{Epifanovsky_21} E.~ Epifanovsky, J. Chem. Phys. {\bf 155}, 084801 (2021).
%===================================== CS reviews =====================================%
\bibitem{Reinhardt_ARPC33_223:1982}
W.~P.~Reinhardt, Annu. Rev. Phys. Chem. {\bf 33}, 223 (1982).
%
\bibitem{Junker_AAMP18_107:1982}
B.~R.~Junker, 
Adv. At. Mol. Phys. {\bf 18}, 207 (1982).
%
\bibitem{Ho_PRC99_1:1983}
Y.~K.~Ho, 
Phys. Rep. C {\bf 99}, 1 (1983).
%
\bibitem{Moiseyev_PR302_211:1998}
N.~Moiseyev,
Phys. Rep. {\bf 302}, 211 (1998).
%
\bibitem{Lindroth_AQC63_247:2012}
E. Lindroth and L. Argenti, 
Adv. Quantum Chem. {\bf 63}, 247 (2012).
%
\bibitem{Maltsev_18} I.~A.~Maltsev, V. ~M. ~Shabaev, R.~V.~Popov,  
               Y.~S.~Kozhedub, G.~Plunien, X.~Ma, and Th.~St\"{o}hlker,
               Phys. Rev A {\bf 98}, 062709 (2018).
%
\bibitem{Maltsev_17} I.~A.~Maltsev, V. ~M. ~Shabaev, I.~I.~Tupitsyn, 
               Y.~S.~Kozhedub, G.~Plunien, 
	        and Th.~St\"{o}hlker, Nucl. Instrum. Methods Phys. Res. Sect. B {\bf 408}, 97 (2017).
%
\bibitem{Simon_79} B.~Simon, Phys. Lett {\bf 71A}, 211 (1979).
%
%
\bibitem{Moiseyev_88} N.~Moiseyev and J.~O.~Hirschfelder, J.~Chem. Phys. {\bf 88},
1063 (1988).
%
\bibitem{Riss_93}
U.~V.~Riss and H.~D.~Meyer, 
J. Phys. B {\bf 26}, 4503 (1993).
%
\bibitem{Elander_98}
N.~ Elander and E.~Yarevsky,  
Phys. Rev. A {\bf 57}, 3119 (1998).
%
\bibitem{Pirkola_22}
P.~Pirkola and M.~Horbatsch,
Phys. Rev. A {\bf 105}, 032814 (2022).
%
\bibitem{LL} L.~D.~Landau and E.~M.~Lifshitz, {\it Quantum Mechanics:
Non-relativistic Theory} (Pergamon, Oxford, 1977).

\end{thebibliography}
\end{document}